\documentclass{elsart-accepted}

\usepackage{natbib}
\bibpunct{(}{)}{;}{a}{}{,}

\usepackage{graphicx}

\usepackage[latin1]{inputenc}

\usepackage{amssymb}

\usepackage{url}

\journal{Deep-Sea Research II}
\date{2004-12-21}

\begin{document}

\begin{frontmatter}


 \title{
       A simple model for the short-time evolution of near-surface
       current and temperature profiles
       }
\author[labelADJ]{Alastair D.~Jenkins\corauthref{cor1}},
 \ead{alastair.jenkins@bjerknes.uib.no}
 \ead[url]{\url{http://www.gfi.uib.no/~jenkins}}
\corauth[cor1]{Corresponding author. Tel.: +47-55 58 2632; fax: +47-55 58 9883.}
\author[labelBW]{Brian Ward}
\address[labelADJ]{Bjerknes Centre for Climate Research, Geophysical Institute,
Allégaten 70, 5007 Bergen, Norway}
\address[labelBW]{Woods Hole Oceanographic Institution,
       Woods Hole, Massachusetts, U.S.A.
}




\begin{abstract}
       A simple analytical/numerical model has been developed for
       computing the evolution, over periods of up to a few hours, of the
       current and temperature profile in the upper layer of the ocean.
       The model is based upon conservation laws for heat and momentum,
       and employs an eddy diffusion parameterisation which is dependent
       on both the wind speed and the wind stress applied at the sea
       surface. Other parameters such as the bulk$-$skin surface
       temperature difference and CO$_2$ flux are determined by application
       of the Molecular Oceanic Boundary Layer Model (MOBLAM) of
       Schlüssel and Soloviev. A similar model, for the current profile
       only, predicts a temporary increase in wave breaking intensity and
       decrease in wave height under conditions where the wind speed
       increases suddenly, such as, for example, during gusts and
       squalls.
       The model results are compared with measurements from the
       lagrangian Skin Depth Experimental Profiler (SkinDeEP)
       surface profiling instrument made during the 1999 MOCE-5 field
       experiment in the waters around Baja California. SkinDeEP made
       repeated profiles of temperature within the upper few metres of
       the water column. 
       Given that no tuning was performed in the model, and that the
       model does not take account of stratification, the results of the
       model runs are in rather good agreement with the observations. The
       model may be suitable as an interface between time-independent
       models of processes very near the surface, and larger-scale
       three-dimensional time-dependent ocean circulation models. A
       straightforward extension of the model should also be suitable for
       making time-dependent computations of gas concentration in the
       near-surface layer of the ocean.
\end{abstract}


\end{frontmatter}

\section{Introduction}
\label{intro}

The temperature of the sea surface, and its relation to that 
of the adjacent atmosphere and water
column, are vitally important parameters for the air--sea exchange of
heat \citep{%
FairallCW-BradleyEF-GodfreyJS-etal:jgr-1996-1295%
} and gas species \citep{%
McNeilCL-MerlivatL,%
FairallCW-HareJE-EdsonJB-McGillisW:blm-2000-63,%
WardB-WanninkhofR-McGillisWR-etal:jgr-2004-xxx%
}.
For global estimation of atmosphere--ocean heat and gas flux, it is necessary
to relate satellite observations of radiative surface skin temperature $T_S$
to the bulk temperature $T_B$ of the upper centimetre or so of the
water column \citep{%
SchluesselP:etal:jgr-1990-13341,%
DonlonCJ-MinnettPJ-etal:jc-2002-353%
}.  This bulk$-$skin temperature difference $\Delta T_{B-S} =
T_B - T_S$ is a function of the ambient
radiative, mechanical, and thermal forcing, and is generally positive and equal
to a few tenths of a degree, primarily as a result of
the cooling of the water by outgoing long-wave radiation emitted in a thin
layer of the order of 1$\mu$m deep. The transition zone from this skin layer
to the bulk is controlled primarily by molecular heat conduction and
has a thickness of the order of 1\,mm.

A number of different
theoretical and numerical models exist for the calculation of 
$\Delta T_{B-S}$ \citep{%
Eifler1992alt,%
Eifler1993,%
Jessup:zapp:hesa:etal1995,%
Soloviev:Schlussel1994%
}.  These models can be quite complex, since they compute many interrelated
parameters, and, since they consider processes with a small time scale, are
generally in time-independent form.  Alternatively, one can
use such techniques as neural network methods to derive
empirically the relation between the forcing parameters and $\Delta T_{B-S}$
\citep{%
WardB:RedfernS:ijrs-1999-3533%
}.

In the upper 10--50\,m layer of the water column, time-dependent
three-di\-men\-sion\-al ocean circulation models may employ simple algebraic
parameterisations for turbulent fluxes of momentum, heat, and mass. They may also
use some form of turbulence closure 
technique \citep[e.g.][]{BurchardH:petersenO:rippethTP:jgr-1998-10453}, albeit
at greater computational expense. Even more
expensive computationally are direct numerical simulations of vortex 
structures \citep[e.g.][]{NagaosaR:pf-1999-1581}.

In this paper we employ a time-dependent model which should be applicable to
the zone intermediate between the surface millimetre layer where the skin
effect occurs and the $\approx$10\,m level which can be resolved by oceanic
circulation models.  The model is thus useful in describing or
parameterising the coupling between the different model scales.  It uses a
simple parameterisation of turbulent flux processes, so that it should be
numerically stable and computationally inexpensive.
The model codes and documentation are specified
in \cite{JenkinsAD-WardB:MTP-2003}.

We compare results of the model with near-surface temperature profiles obtained
using the Skin Depth Experimental
Profiler \citep[SkinDeEP,][]{WardB-WanninkhofR-etal:jaot-2004-207}. 
It is an autonomous profiler, carrying high-resolution
temperature sensors to provide a record of the bulk
temperature. It was deployed during the Marine Optical Characterisation
Experiment (MOCE-5), cruise in the Gulf of California during the period 
1999 October 1--21 \citep{WardB-MinnettP:jgr-2004b-xxx}.

\section{Model formulation and specification}

\subsection{Physical basis of the model}
The flux of momentum, heat, and mass between the atmosphere and the ocean is to
a large extent controlled by turbulence in the atmospheric and oceanic boundary
layers, although stratification \citep{MoninAS-ObukhovAM:tgian-1954-163}, 
the Earth's rotation \citep{EkmanVW:amaf-1905-2}, surface waves~\citep{%
Janssen1989,%
Jenkins19871,%
Jenkins1989n,%
Weber19831%
}, 
the
presence of laminar layers close to the surface, and the short-wave and long-wave radiation
balance \citep{SchluesselP:etal:jgr-1990-13341} 
can also be important.  The effect of turbulence may be
modelled by eddy viscosity \citep{MadsenOS:jpo-1977-248} 
or turbulence closure models 
which have reached a great degree of complexity and
sophistication \citep[e.g.][]{BurchardH:petersenO:rippethTP:jgr-1998-10453}.  
However, one difficulty that may arise when applying
turbulence closure models in calculating properties very close to the sea
surface is the choice of boundary condition: in general, it may be necessary to
specify a roughness length in a more-or-less arbitrary fashion, and the values
of mean velocity, temperature, and 
concentration near the water surface may have a
behaviour which is close to singular and thus difficult to resolve numerically.
The
partial differential equations which are obtained from a turbulence closure
formulation may be rather complex and their solutions may pose analytical and/or
numerical difficulties.

The model approach employed in this paper is designed to avoid some of the
difficulties described above, by combining physical conservation laws with some
general properties of turbulent boundary layers and empirical
observations.  The physical conservation laws are those of momentum, heat, and
mass: the rate of change of the momentum of the water column is given by the
applied wind stress; likewise the atmosphere--ocean heat flux is given by (or
determines) the rate of change of the total heat content of the water column, 
and the flux of a
substance through the sea surface determines the rate of change of its
concentration in the water column.  The empirical observations are as follows:
the mean velocity (current) at the sea surface is determined, not by a
relationship between the applied wind stress and the vertical velocity
gradient, mediated by a viscosity or turbulent eddy viscosity, but, instead, as
a fraction $\lambda\approx0.02$ of the wind speed~$U$.  Thus, the value of 
the wind-induced surface current has a
relatively stable behaviour, which is in agreement with the practical
engineering calculations made in connection with observations of the drift of
floating objects and of oil slicks, and the measurement of near-surface
currents using moored current meters and surface drifters \citep{%
AudunsonT:osc-1979,%
Huang1979,%
HuangJC:OSC-1983-313,%
Jenkins19871,%
Jenkins198725%
}.  Although the physical reason for this
behaviour of the surface current may not be strictly evident, we may consider
that an increasing wind stress will cause an increasing vertical velocity
shear, but it will also cause an increase in turbulent motion which will
transport momentum more rapidly downwards, thus decreasing the shear. 
The corresponding conditions for temperature and concentration would be equal
temperatures and concentrations just above and below the water surface,
although corrections need to be applied for the presence of a laminar boundary
layer and the thermal radiation balance \citep{SchluesselP:etal:jgr-1990-13341}.

The effect of turbulence in transporting momentum, heat, and mass vertically
can 
be parameterised in terms of a mixing length proportional to the distance from
the surface, as in the usual turbulent boundary-layer of Prandtl and
von~Kármán.
However, we show that if we assume a simple 
functional form (Eq.~\ref{eq-uzt} below) for the
time-dependent behaviour of the current, temperature, etc., combined with the
conservation laws for momentum, heat, and mass, and the empirical surface
boundary conditions discussed above, an effective `mixing length'
scale
appears as a consequence.  Although the physical boundary conditions at the sea
surface and at the sea bottom are fundamentally different, 
there is evidence from observations of wind-induced 
surface currents
\citep{%
Huang1979,%
HuangJC:OSC-1983-313,%
JenkinsAD:CMO-1984-xxx,%
JenkinsAD-OlsenRB-ChristianidisS:CCM-1986-20%
},
temperature 
(\citealp{SoetjeKC-HuberK:mfeA-1980-69%
}; \citealp[][section~7.3.2]{%
BurchardH:atmmw-2002%
}),
and the behaviour of air bubbles in the water
column~\citep{ThorpeSA:jpo-1984-855}, 
that a substantial vertical velocity shear can exist near the surface,
consistent with an increase in 
effective eddy
viscosity with increasing distance from the
surface, as in Madsen's~\citeyear{MadsenOS:jpo-1977-248} approach.

\subsection{Predicted current profile}
\label{sec-pcp}

For an ocean initially at rest, we consider at first the following
evolution of the current $u(z,t)$ at time $t$ and depth $-z$:
\begin{equation}
u(z,t)=\lambda U\exp \left[\lambda Uz/({u_{*}}^{2}t)\right],  \label{eq-uzt}
\end{equation}
where $u_*$ is the friction velocity within the water column, $\tau = \rho_w
{u_*}^2 = \rho_a{U_*}^2$ being the wind stress. $U_*$, $\rho_a$, and $\rho_w$
are the friction velocity in the atmosphere, the air density and the water
density respectively.
Equation \ref{eq-uzt} satisfies the conservation of momentum:
\begin{equation}
\rho_w{d \over dt}\int_{-\infty}^0 u(z,t)\,dz = \rho_a {U_*}^2,
\end{equation}
where for simplicity we neglect the Coriolis force.

The current $u(z,t)$ in (\ref{eq-uzt}) obeys the equation
\begin{equation}
\frac{\partial u}{\partial t}=-\frac{{u_{*}}^{2}z}{\lambda U}\, \,
\frac{\partial ^{2}u}{\partial
z^{2}}, \label{eq-pde}
\end{equation}
with $-{u_{*}}^{2}z/(\lambda U)$ playing a role similar to that 
of an eddy viscosity,
as in turbulent boundary layer flow near a rigid wall.  We may also consider
this `quasi eddy viscosity' to be a product of a turbulent velocity scale
equal to $u_*$ and an effective turbulent roughness length equal 
to $-u_* z/(\lambda U)$. 

If the
wind speed is allowed to vary with time we can alter (\ref{eq-uzt}) so that it
becomes
\begin{equation}
u(z,t) = \int_0^t \lambda {dU(t')\over dt'} \exp\left[{\lambda U(t')\,z \over
\left(u_{*}(t')\right)^{2}\,(t-t')}\right] dt'.
\end{equation}
In this case, the partial differential equation (\ref{eq-pde}) does not hold,
as the system retains a `memory' of the wind forcing at previous points in
time.  The memory is effectively longer at large depths than near the surface.

\begin{figure}
\includegraphics[width=1.1\textwidth,bbllx=119,bblly=74,bburx=386,bbury=258]{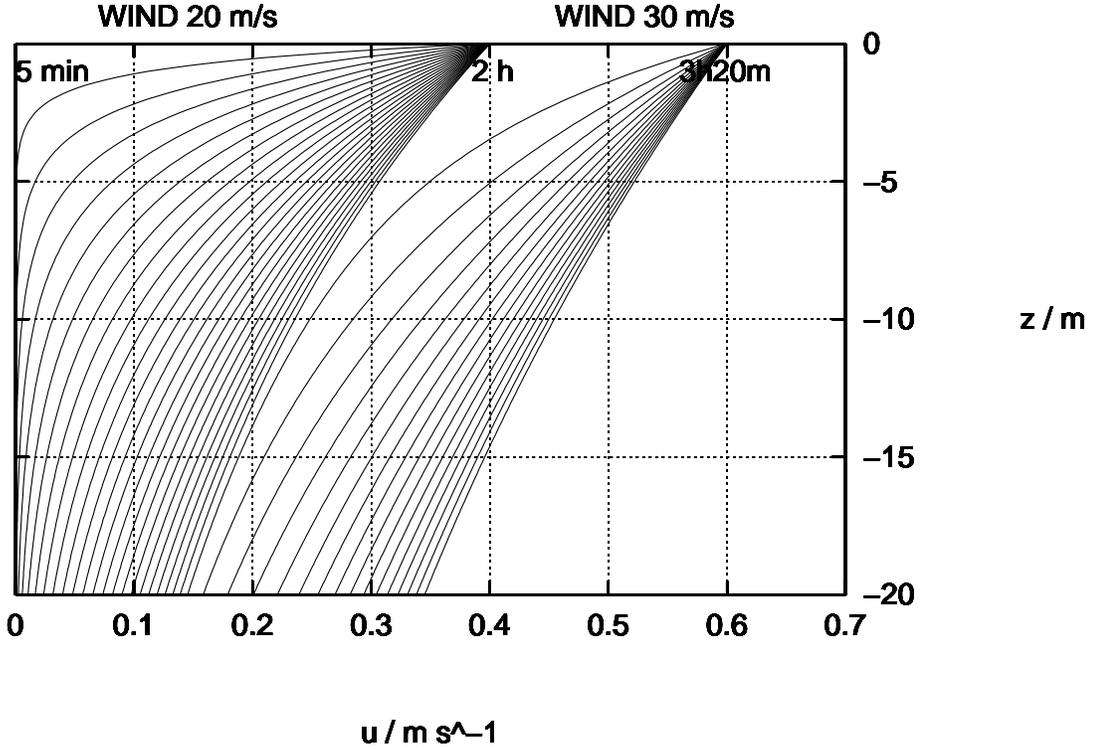}
\caption{\label{figcur}
Example of the evolution of a current profile
using the present model. \protect\cite[From][]{JenkinsAD:waves-2001-494}.
\copyright~2001, ASCE.
Reproduced by permission of the publisher, ASCE.
\protect\url{http://www.pubs.asce.org}
}
\end{figure}

Figure~\ref{figcur}, from \citet{JenkinsAD:waves-2001-494}, 
shows the model predictions for the  current induced by a wind of 20 m/s blowing
for 2 hours followed by an increased wind of 30 m/s. The current profiles are
calculated every 5 minutes. We assume that $\lambda = \rm 0.02$, and $u_* =
\left({C_D\rho_a/\rho_w}\right)^{1/2}U$, with $\rho_a/\rho_w = 1/800$,
$C_D = A + BU$, $A=0.8\times10^{-3}$, and
$B=0.065\times10^{-3}\,{\rm m}^{-1}{\rm s}$ \citep{WuJ:jgr-1982-9704}.

{\em Effect of the modelled current on surface waves}\par
\begin{figure}
\includegraphics[width=1.1\textwidth,bbllx=119,bblly=84,bburx=386,bbury=228]%
{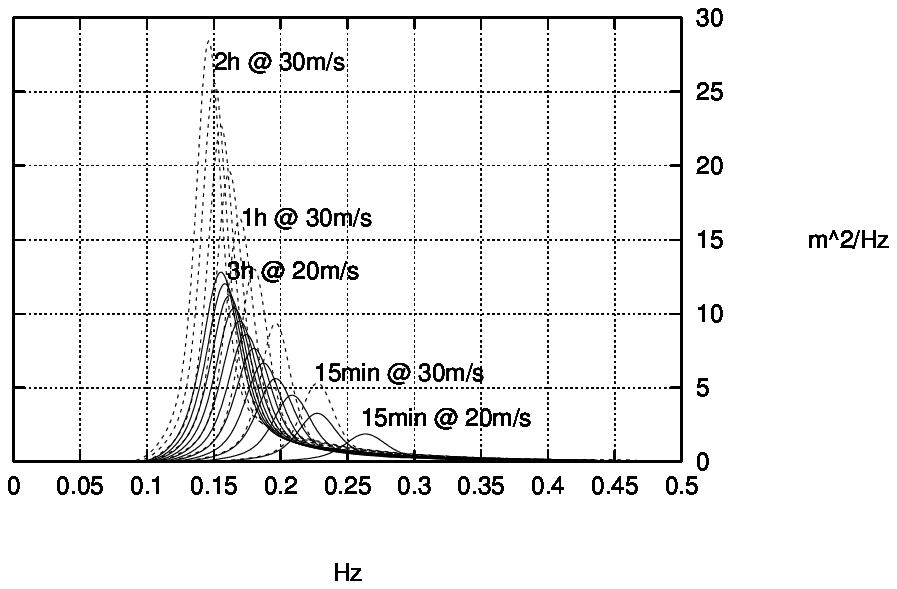}\\
\includegraphics[width=1.1\textwidth,bbllx=119,bblly=84,bburx=386,bbury=228]%
{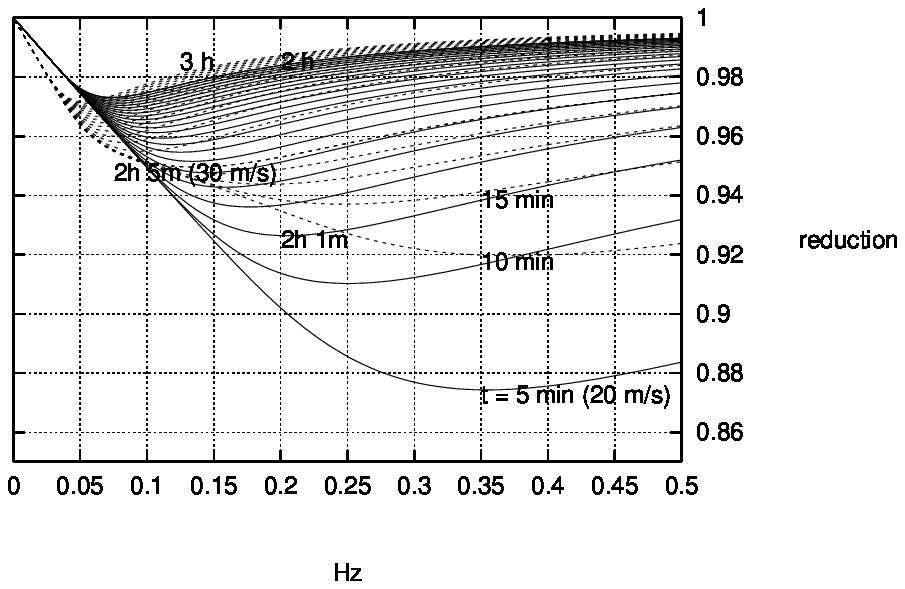}\\
\caption{\label{fig-wavered}
Wave spectra from the Donelan/Hamilton/Hui model, and
  computed reduction in amplitude of maximum non-breaking waves.
\protect\cite[From][]{JenkinsAD:waves-2001-494}. \copyright~2001, ASCE.
Reproduced by permission of the publisher, ASCE.
\protect\url{http://www.pubs.asce.org}.
}
\end{figure}

Wave spectra have been calculated by using
the \citet{Donelan:Hami:Hui1985}
formulation for limited fetch, transformed by means of the wave group velocity
to apply to winds blowing for limited times. Calculations have also been made
of the effect of the computed vertical current shear on the wave height
required
for wave breaking, according to the theory of
\citet{BannerandPhillips1974}. Results of
these calculations (Figure~\ref{fig-wavered}) 
indicate that a rapid increase in wind speed will tend to
increase the intensity of wave breaking so that the wave height will decrease
by a small factor, of order 6\% according to linear wave theory, but this
factor
may be as much as 25\% according to time-dependent nonlinear numerical
simulations
by \citet{BannerM:BabaninAV:YoungIR:jpo-2000-3145}.

\subsection{Heat flux}

If we now consider the diffusion of heat within the water column, we obtain a
corresponding formula for the evolution of the temperature $T$:
\begin{equation}
T(z,t) = T_0 + \int_0^t {dT(0^-,t')\over dt'}
\exp\left[{\lambda U(t')\,z
\over \left(u_{*}(t')\right)^{2}\,(t-t')}\right] dt', \label{eq-model-T}
\end{equation}
where $T_0$ is the initial temperature, and
$T(0^-,t)$ is the bulk temperature just below the surface skin layer.
Here we assume that turbulence diffuses heat at the same rate as it diffuses
momentum. We may adjust the relative values of the turbulent
diffusion coefficients for heat and momentum 
by altering the value of the parameter $\lambda$.  The model
is assumed to be horizontally homogeneous, and the effect of density
stratification on the turbulent diffusion of heat is neglected.

\subsection{Characteristic behaviour of the model}
\label{sec-cbm}
It may be considered that the model presented in this paper produces results
which are at variance with the turbulence closure and eddy diffusion models
normally employed.  In particular, the evolution of the current, temperature,
or concentration of a tracer, driven by a step-function forcing of wind stress,
surface temperature, etc., produces results which imply that the eddy viscosity
or diffusivity increases with time and tends to an infinite value for infinite
time.  However, this behaviour does not make the model invalid.  We are here
assuming infinite depth and neglecting the influence of stratification and
rotation, and in such a case there is no natural limit to the size of turbulent
eddies which may be produced.  Such eddies may grow indefinitely, and the
effective eddy viscosity or diffusivity has a term which increases linearly
with time.  If the eddy viscosity may be
regarded as proportional to the product of a velocity scale ($u_*$) and a
length scale (the size of turbulent eddies), this will mean that such
eddies should increase in size linearly with time, consistent with the lack of
a natural limit to eddy size.  For a constant wind speed $U$ and surface bulk
temperature $T(0^-)$, a steady state would have a constant current $\lambda U$
and a depth-independent temperature $T(0^-)$. However, this steady state will
be reached very slowly in the assumed case of infinite water depth and zero
Coriolis force.

In addition, it is by no means assured that the vertical transport may
be parameterised by means of an eddy viscosity or diffusivity.  The model may
be criticised for 
the fact that the velocity, temperature,
or concentration gradient decreases to zero with increasing time: however,
it is well known that it is not uncommon in turbulent flows for 
counter-gradient transport to occur, 
so that a finite flux of momentum/heat/mass under
zero-gradient conditions should not be an unusual effect.  It is common
in the modelling of atmospheric turbulent diffusion for non-local
transport algorithms to be used \citep[e.g.][]{StullRB:blm-1993-21}, 
and such algorithms almost invariably
lead to counter-gradient transport in at least part of the domain, in agreement
with results using large-eddy simulation algorithms
 \citep{Skyllingstad:Denbo1995}.

It is in fact 
possible to re-write Eq.~\ref{eq-pde} as an advection--diffusion equation,\footnote{H.~Burchard, personal
communication.} 
\[
{\partial u\over\partial t} + w{\partial u\over\partial z} -
{\partial\over\partial z}\left(\nu_E{\partial u\over\partial z}\right) = 0,
\]
employing the same value for eddy viscosity, $\nu_E = -u_*^2z/(\lambda U)$,
 as mentioned
previously.  The turbulent flux of momentum is thus expressed here as the sum
of a downgradient and a no-gradient flux.  The no-gradient flux is described 
by a downward vertical advection velocity 
$w=-u_*^2/(\lambda U)$.  Note that this advection velocity is in general much
smaller than $u_*$, so in our opinion it should not be regarded as unphysical.

The eddy viscosity may still be described as
the product of a velocity
scale $u_*$ and a turbulent length scale $-u_*z/(\lambda U) =
-\lambda^{-1}\left({C_D\rho_a/\rho_w}\right)^{1/2}z$.  If we let the eddy viscosity profile
be described in terms of a von~Kármán `constant' $\kappa$, i.e.\ 
$\nu_E = -\kappa u_* z$, we have $\kappa = u_*/(\lambda U) =
\lambda^{-1}\left({C_D\rho_a/\rho_w}\right)^{1/2}$.  This value of $\kappa$ is generally
smaller than the `classical' value of $\approx0.4$: for the parameter values
used here (see section~\ref{sec-pcp}), $\kappa$ increases with wind speed,
being 0.050 at $U=0$\,m/s and 0.081 at $U=20$\,m/s.  

\section{Measurements}

A detailed description of the SkinDeEP instrument,
which uses an FP07 thermistor and a high-resolution platinum wire  (Pt)
temperature
sensor, is given by \citet{wardB:minnettP:GTWS-2002-167} and
\citet{WardB-WanninkhofR-etal:jaot-2004-207}.
The instrument is able to rise and sink autonomously, changing its
density by inflating and
deflating an external neoprene sleeve, and measurements are made during its
ascending phase only.

The instrument was deployed during the MOCE-5 cruise, in the
waters around Baja California \citep{WardB-MinnettP:jgr-2004b-xxx}.  
Data were acquired from
ten stations: Table~\ref{tab-stat} summarises the deployment information for
the three stations where the results are shown in this report.
The measurements reported here are from the
FP07 thermistor, as data from the Pt sensor were not available for this cruise.

\begin{table}
\caption{\label{tab-stat} SkinDeEP deployment information for the measurements
presented in this paper}\par\smallskip
{\footnotesize
\begin{tabular}{|c|c|c|c|l|c|}\hline
Date&Stat.&Station&Times&No.~of&Position\\
    &no.  &name   &LST  &profiles&      \\ \hline
1999-10-04&\hfill4&Punta Magdalena\hfill&10:59--13:24
  &\hfill120&25$^\circ$\,09.49$'$N, 112$^\circ$\,59.52$'$W\\
1999-10-10&\hfill10&\hfilneg T/S Irwin\hfill&11:12--14:13&\hfill161
           &22$^\circ$\,31.48$'$N, 109$^\circ$\,35.43$'$W\\
1999-10-14&\hfill14&Isla San Esteban 2\hfill&06:15--07:44&\hfill78
           &28$^\circ$\,36.15$'$N, 112$^\circ$\,32.04$'$W\\ \hline
\end{tabular}
}
\end{table}

Other observations made include the following:
\begin{itemize}
  \item Sea-surface skin temperature by the M-AERI passive infrared radiometric
interferometer, using the 500--3000\,cm$^{-1}$ wavelength range, which has an
accuracy of better than 0.05$^\circ$ \citep{%
minnettPJ:etal:jaot-2001-994,%
minnettPJ:wardB:ENVS-2000-xxx%
};
  \item Bulk water temperature measurements from the ship intake and from a
floating sensor (an inverted hard plastic helmet filled with foam with a
thermistor just below the waterline);
  \item Air temperature, wind speed, and net heat flux.
\end{itemize}

All three measurement series presented here
are for periods with rather light winds, between 0.7 and
4.5~m\,s$^{-1}$. The air temperature is less than the sea surface temperature
(measured by SkinDeEP) on October 4, is greater than the surface temperature on
October 14, and on October 10 the air and sea surface temperatures are quite
similar.  The net heat flux is positive (from air to sea) in all three cases.
The net heat flux is primarily incident short-wave radiation, and is
mostly absorbed in the upper 0.1\,m of the water
column \citep{SchluesselP:etal:jgr-1990-13341}.

Shipboard logistics dictated that SkinDeEP was deployed during the
afternoon to facilitate other operations, which is the timeframe when
measurements were made on October 4 and 10. An exception to this routine was
made on one day (October 14), when the profiler was deployed 
early in the morning. 
As measurements began shortly before sunrise, this was
the only nighttime dataset,
and was too limited to provide any
conclusive evidence for daytime--nighttime differences.

\section{Model results}

\begin{figure}
{\centering \resizebox*{1\columnwidth}{!}{\rotatebox{-90}%
{\includegraphics{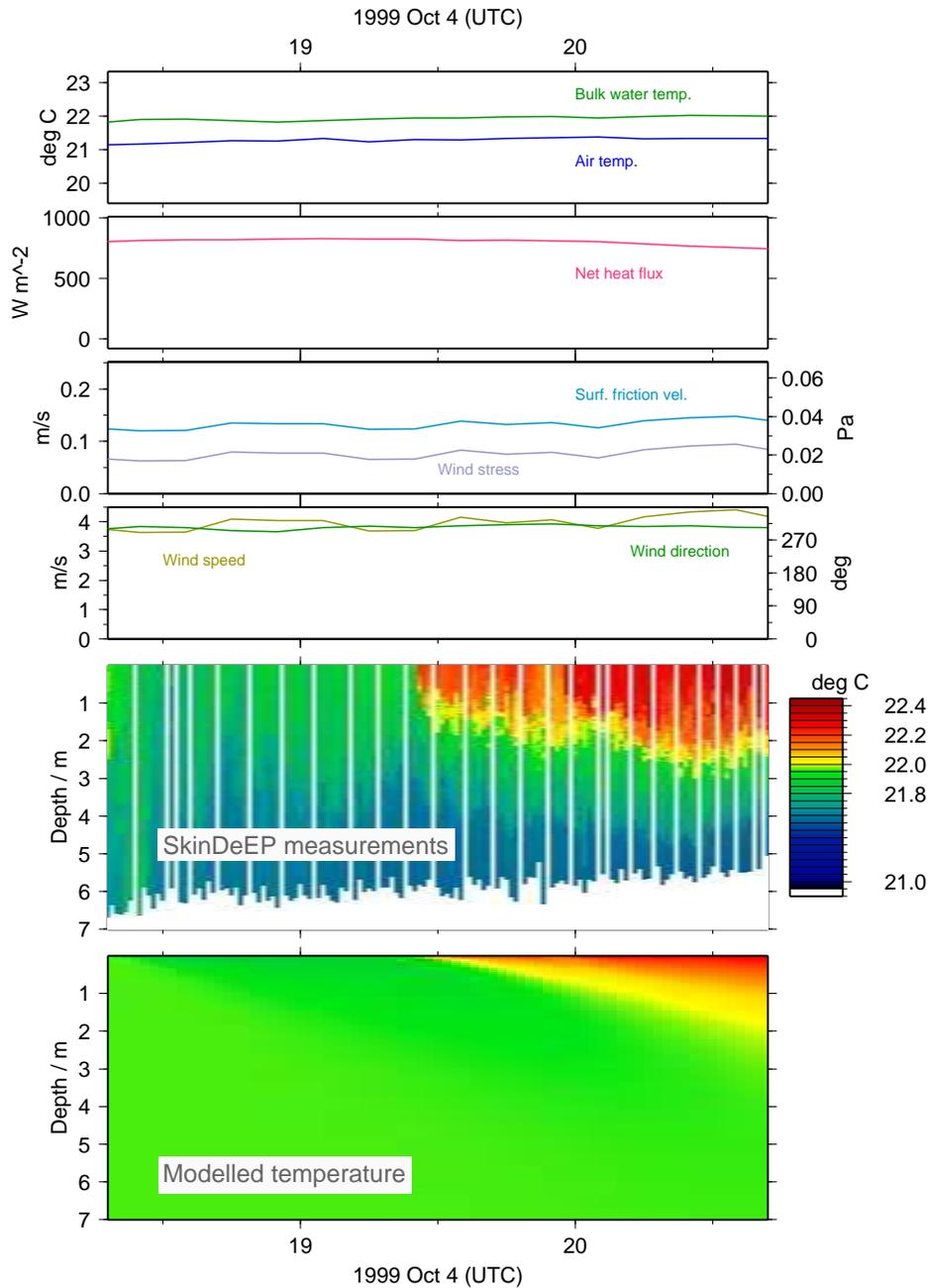}}} \par}
\caption{\label{figoct04} Observations and model results from Station Punta
Magdalena, 1999~October~4.
}
\end{figure}

\begin{figure}
\rule{0pt}{0pt}\\[-10mm]
{\centering \resizebox*{0.65025\columnwidth}{!}{\rotatebox{0}%
{\includegraphics{oct10_final.eps}}}\\
\rule{27mm}{0pt}\includegraphics[width=0.55379\columnwidth]{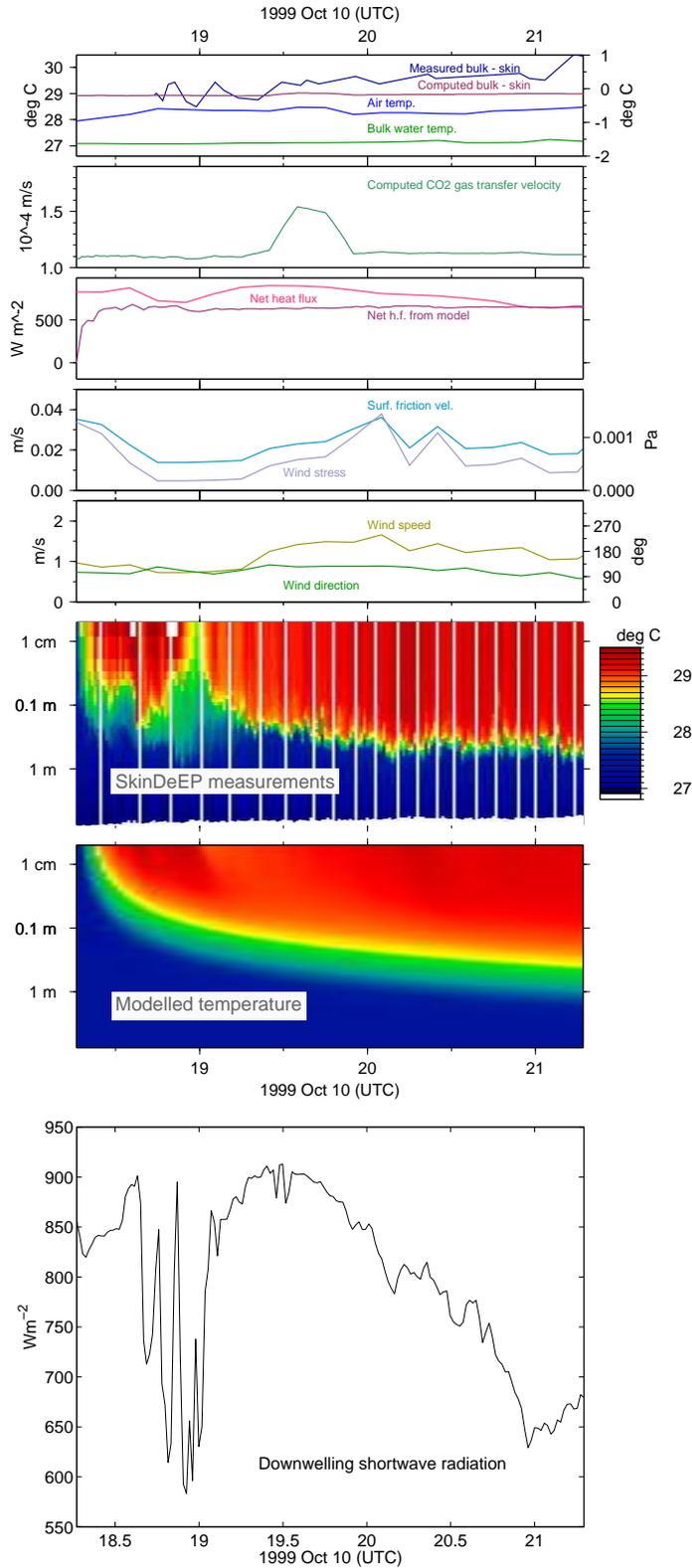}}%
\\[-9.5mm]
\caption{\label{figoct10} 
\protect\footnotesize
Observations and model results from Station T/S
Irwin, 1999 October 10, including details of the downwelling shortwave
radiation (bottom frame). The modelled net heat flux is obtained from
Eq.~\protect\ref{eq-mod-heatflux}. 
The modelled  bulk$-$skin temperature difference and
CO$_2$ flux are computed from the
observed meteorological parameters from the ship
(wind speed, air temperature, humidity, heat fluxes, precipitation)
and uppermost water temperature measured by
SkinDeEP, using the MOBLAM model of P.~Schl\"ussel and A.~V. Soloviev.
}
\end{figure}

\begin{figure}
\resizebox*{1\columnwidth}{!}{\rotatebox{-90}%
{\includegraphics{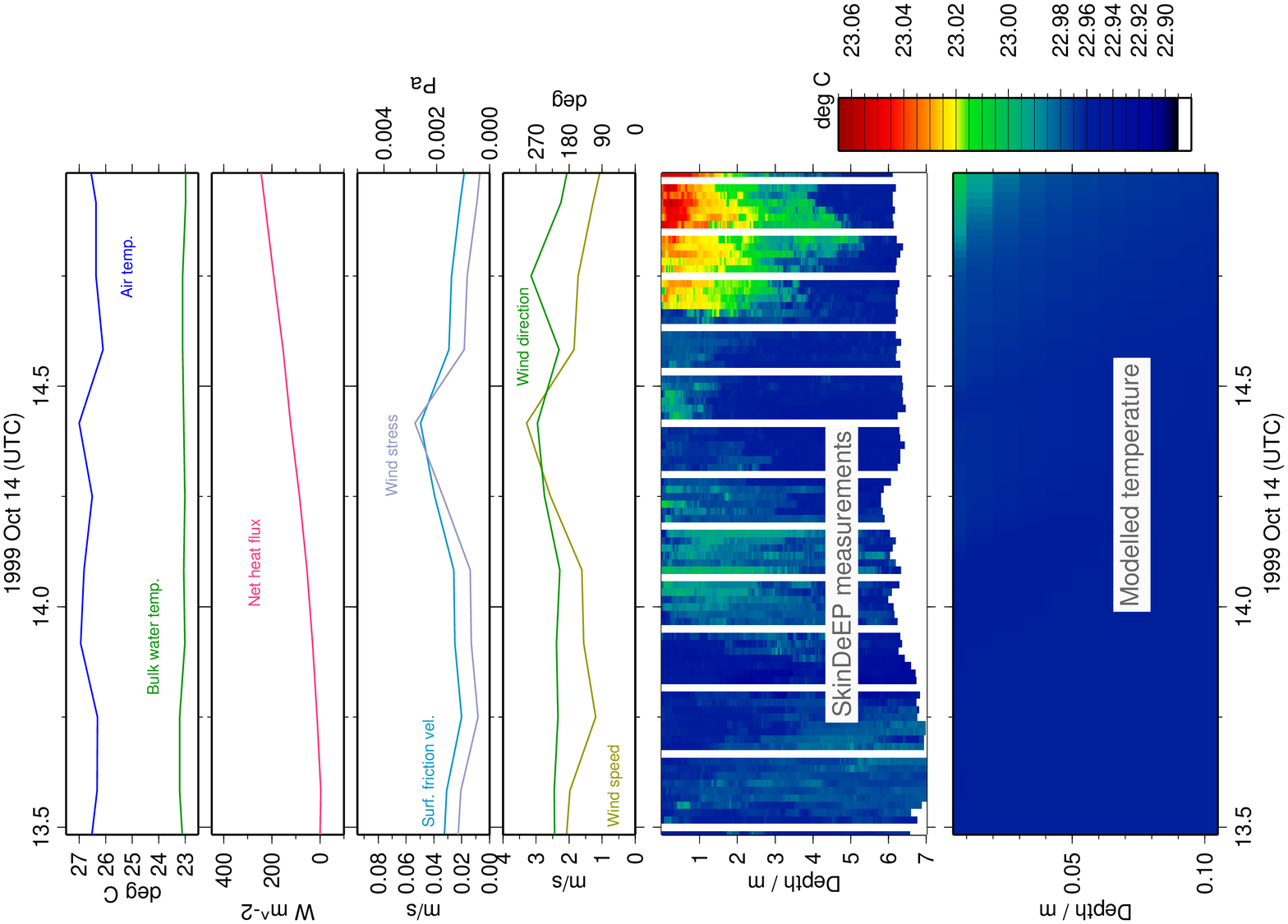}}}
\caption{\label{figoct14} Observations and model results from Station Isla San
Esteban 2, 1999 October 14}
\end{figure}

Figures~\ref{figoct04}--\ref{figoct14} show SkinDeEP measurements from three of
the MOCE-5 stations, together with background atmospheric parameters and net
heat flux.  Results of model simulations from equation \ref{eq-model-T} are
shown at the bottom of each figure.  The model is started with a uniform
temperature at the time
corresponding to the left-hand side of the plot, and the surface boundary
condition $T(0^-,t)$ is set equal to the uppermost water temperature 
measured by the SkinDeEP profiler
(at a depth of approximately 0.5\,cm, in the bulk, below the surface skin 
layer).  The value $u_*= \left({C_D\rho_a/\rho_w}\right)^{1/2}U$ 
in (\ref{eq-model-T}) is 
computed using the
drag coefficient formula of \cite{WuJ:jgr-1982-9704}:  $C_D = A + BU$,
$A=0.8\times10^{-3}$, and
$B=0.065\times10^{-3}\,{\rm m}^{-1}{\rm s}$.

 The ocean is assumed to be horizontally homogeneous,
and neither the effect of stratification nor of the Coriolis force are taken
into account.  The current and waves were not calculated in this case,
since no observations of current or waves were available.

{\em 1999 October 4:}\par
In this situation the air temperature is lower than the sea temperature, but
the net positive heat flux should tend to increase the bulk temperature near
the sea surface.  The observed warming of the surface layer at around 12:30 is
likely to be due to horizontal advection.  Nevertheless, the model does predict
an increase in depth of the warm surface layer ($T>22^\circ$C) to 2\,m by the
end of the observation period, which is of the same order as the depth increase
which is actually observed.  It may also be the case that the model
parameterization,
which predicts a `von Kármán coefficient' lower than the classical value of
0.4, is underpredicting the amount of vertical mixing.  In such a case, further
investigations are required to determine how the model parameters may be
tuned.

{\em 1999 October 10:}\par
Here the modelled increase in the warm surface layer depth during the
observation period is quite similar to that actually observed.  The drop in the
observed surface temperature at 19:00 may be associated with the decrease in
net heat flux in the previous 15 minutes, but the associated changes in the
depth of the warm surface layer are not reproduced in the model.  The greater
wind speed in the period around 20:00, together with the positive air$-$sea
temperature difference, tend to increase the temperature in the upper few
centimetres of the ocean, and this is reproduced in both measurements and model
results. Again, the agreement between model results and measurements is good,
considering the fact that the model does not take stratification into account.

The sudden cooling in the water surface at about 19:00 on October 10
coincided with the passage of a cloud, which is visible in the
downwelling shortwave data (bottom frame of Figure~\ref{figoct10}). 
This had a dramatic and
immediate effect on the surface warming, presumably because of the
absence of wind mixing.

{\em 1999 October 14:}\par
In this case the model appears to make a poor prediction of the temperature
evolution of the surface layer of the ocean, when compared with the
measurements.  However, the temperature in the water column is
rather uniform, varying by less than 0.15\,K over the whole domain. The
turbulent diffusion predicted by the model suppresses any developing
non-uniformity in the temperature profile,
an effect which would be mitigated if stratification were taken into account.

\subsection*{Additional measurements and model results}
For the October 10 data set, a number of additional measurements and model
results are presented.  The downward heat flux $Q_m(t)$ through the surface
computed from the present model is given by:
\begin{equation}
Q_m(t) = {\rho_w C} \int_0^t {dT(0^-,t')\over dt' }
\left[{\left(u_{*}(t')\right)^{2}
\over \lambda U(t')}\right]\,dt', \label{eq-mod-heatflux}
\end{equation}
with $T(0^-,t')$ being the uppermost water temperature measured by SkinDeEP.

The bulk$-$skin temperature difference and gas flux are computed from the
observed meteorological parameters from the ship
(wind speed, air temperature, humidity, heat fluxes, precipitation)
and uppermost (bulk) water temperature measured by
SkinDeEP,
using the Molecular Oceanic Boundary LAyer Model (MOBLAM)%
\footnote{MOBLAM is available on the Internet from
\url{http://www.nova.edu/ocean/gasex/Moblam8.f90}, the version used here
being from 1999 February 10. The model is described
on \url{http://www.nova.edu/ocean/gasex/micro.html}.}, by Peter Schl\"ussel and
A.~V.~Soloviev.
The MOBLAM model is a surface renewal model, which takes account of solar and
long wave
radiation, turbulent and diffusive
fluxes of sensible and latent heat, and impact of
rain on the cool skin \citep{%
SolovievAV:Schlussel:blm-1996-45,%
SchluesselP:Soloviev:Emery:blm-82-437,%
CraeyeC:Schluessel:blm-1998-349%
}.

{\em Heat flux:}\par
The heat flux computed by the present model, after a short period of
adjustment, settles down to a fairly constant level which is roughly equal to
the net heat flux deduced from observations.  Given the simplicity of the model
and the fact that no parameter adjustment has been made, this is remarkably
good agreement.

{\em Bulk$-$skin temperature difference:}\par
The bulk$-$skin temperature difference calculated from the MOBLAM model
gives values of between $-$0.2\,K and $-$0.15\,K, which are considerably
smaller than the measured values using SkinDeEP and the M-AERI interferometer.
This considerable discrepancy requires further investigation.

{\em Gas flux:}\par
The CO$_2$ gas transfer velocity, computed by MOBLAM from the
observed meteorological parameters from the ship
(wind speed, air temperature, humidity, heat fluxes, precipitation)
and uppermost (bulk) water temperature measured by
SkinDeEP, is
fairly constant, except for enhanced values between 19:20 and 20:00.  The
increasing values are associated by an increase in wind stress, which will
increase the amount of turbulent diffusion, whereas the
decreasing values are associated with an increase in the sea surface
temperature and consequent decrease in CO$_2$ solubility.

{\em Daytime--night-time differences:}\par
Shipboard logistics dictated that SkinDeEP was deployed during the
afternoon. The one exception was on October 14, but the data set was too
limited to provide any conclusive evidence for daytime--night-time differences.

\section{Discussion and conclusion}
The time-dependent model presented in this paper is designed to provide an
economical method of computing the evolution of the
current (momentum), temperature, heat and mass flux in
the near-surface layer of the ocean. Although
it does not take account of stratification or of the Earth's rotation,
nevertheless, given the simplicity of the model and the fact that no tuning
was performed on the model parameters, it is shown to provide useful results
for the time evolution of the temperature distribution in the oceanic surface
layer over the measurement periods of the order of 2~hours.
Some of the discrepancies we found may be explained as the result of horizontal
advection, and by the model not taking account of the effect of
stratification.  It may also be the case that the amount of vertical mixing is
underestimated, as indicated to a certain extent by the observations of
Figure~\ref{figoct04}, and by the low von~Kármán `constant' of the model,
discussed in section~\ref{sec-cbm}.  It should be noted, however, that the
turbulent flow and mixing 
in the near-surface boundary layer of the ocean need not be
represented by the
`classical' von~Kármán `constant' value of 0.4, as the physical conditions are
by no means the same as those in a classical turbulent wall layer.  
A resolution of the problem of turbulent mixing intensity
may be obtainable in the future by conducting an analysis of 
further, more detailed time-dependent coupled
observations of velocity, temperature, and turbulence in the 
near-surface atmospheric and oceanic boundary layers. 

The computed net surface heat flux values are largely consistent
with observed values.  A suitable application for the model would be as an
interface between complex time-independent models for interfacial flux of
momentum, heat, and mass, and time-dependent three-dimensional numerical
models of ocean circulation which employ turbulence closure schemes and
account for stratification and the Coriolis force.  

The flux of CO$_2$ and other gas species through the sea surface can be
computed using MOBLAM or similar time-independent surface models.  A
straight\-forward extension of the present time-dependent model should be
capable of extending these flux predictions and computing the concentration and
flux of CO$_2$ further into the water column.  

It should be possible to
test the present model against time-dependent models with one (vertical)
spatial dimension, such as the warm layer model of
\cite{PriceJF-WellerRA-PinkelR:jgr-1986-8411}, and with models which predict
bulk fluxes such as the TOGA--COARE algorithm \citep{%
FairallCW-BradleyEF-GodfreyJS-etal:jgr-1996-1295,%
FairallCW-BradleyEF-RogersDP-etal:jgr-1996-3747%
}\footnote{The algorithm is currently available on the Internet at URL
\url{http://www.coaps.fsu.edu/COARE/flux_algor/flux.html}}.
Such a study, including the use of turbulent closure models for stratified
flows, and a more detailed evaluation of the effect of horizontal
advection, would be a suitable topic for a future paper.

\subsection*{Acknowledgements}

The work was funded by Research Council of
Norway projects 127872/720 and 155923/700, by the Norwegian high performance
computing consortium under grant NN2932K, and by European
Commission Contract No.~ERBFMBICT983162.
Additional funding was provided by NSF grant OCE-0326814.
Observational
data are from the Marine Optical Characterisation
Experiment (MOCE-5), conducted by the U.S. National Aeronautic
and Space Administration and National Oceanic and Atmospheric Administration.
The computer code MOBLAM was made available on the Internet at address
\url{http://www.nova.edu/ocean/gasex/Moblam8.f90} by Dr.~Peter Schl\"ussel.
The authors would like to thank Walter Eifler, Hans Burchard, and one anonymous
referee for constructive criticism of the manuscript.
This is Publication No.~A~84 from the Bjerknes Centre
for Climate Research, and Woods Hole Oceanographic Institution
Contribution No.~11347.

\end{document}